# The effect of the misfit dislocation on the in-plane shear response of the ferrite/cementite interface

Jaemin Kim[a], Keonwook Kang[b,**], Seunghwa Ryu[a,*]

[a] Department of Mechanical Engineering & KI for the NanoCentury, Korea Advanced Institute of Science and Technology, Daejeon 34141, Republic of Korea

[b] Department of Mechanical Engineering, Yonsei University, Seoul 03722, Republic of Korea

[*] Corresponding author. Tel.: +82-42-350-3019; fax: +82-42-350-3059.

Email address: ryush@kaist.ac.kr (Seunghwa Ryu)

[**] Co-corresponding author. Tel.: +82-2-2123-2825; fax: +82-2-312-2159.

Email address: kwkang75@yonsei.ac.kr (Keonwook Kang)

## Abstract

Although the pearlitic steel is one of the most extensively studied materials, there are still questions unanswered about the interface in the lamellar structure. In particular, to deepen the understanding of the mechanical behavior of pearlitic steel with fine lamellar structure, it is essential to reveal the structure-property relationship of the ferrite/cementite interface (FCI). In this study, we analyzed the in-plane shear deformation of the FCI using atomistic simulation combined with extended atomically informed Frank-Bilby (xAIFB) method and disregistry analyses. In the atomistic simulation, we applied in-plane shear stress along twelve different directions to the ferrite/cementite bilayer for Isaichev (IS), Near Bagaryatsky (Near BA) and Near Pitsch-Petch (Near PP) orientation relationship (OR), respectively. The simulation results reveal that IS and Near BA ORs show dislocation-mediated plasticity except two directions, while Near PP OR shows mode II (in-plane shear) fracture at the FCI along all directions. Based on the xAIFB and disregistry analysis results, we conclude that the in-plane shear behavior of the FCI is governed by the magnitude of Burgers vector and core-width of misfit dislocations.

## 1. Introduction

Pearlitic steel has been used in many engineering applications such as bridge cable, rail steel and tire cord for its high strength and good ductility. Pearlitic steel has a three-dimensionally interconnected lamellar structure composed of ferrite (α-Fe, body centered-cubic structure) and cementite ($Fe_3C$, orthorhombic structure) phases with specific orientation relationships (ORs) (Hillert 1962, Zhang, Esling et al. 2007). Extensive studies (Embury and Fisher 1966, Langford 1970, Cooke and Beevers 1974, Hyzak and Bernstein 1976, Langford 1977, Taylor, Warin et al. 1977, Sevillano 1991, Embury and Hirth 1994, Bae, Nam et al. 1996, Janecek, Louchet et al. 2000, Hono, Ohnuma et al. 2001, Hohenwarter, Taylor et al. 2010, Li, Choi et al. 2011, Zhang, Godfrey et al. 2011, Li, Yip et al. 2013, Baek, Hwang et al. 2014, Iacoviello, Cocco et al. 2015, Hohenwarter, Volker et al. 2016, Kapp, Hohenwarter et al. 2016) on the pearlitic microstructure have been conducted to explain its unique mechanical properties. In these works, various roles of the ferrite/cementite interface (FCI) were reported such as a dislocation barrier (Langford 1970, Hyzak and Bernstein 1976, Langford 1977, Kapp, Hohenwarter et al. 2016) and nucleation site (Sevillano 1991, Embury and Hirth 1994, Janecek, Louchet et al. 2000, Guziewski, Coleman et al. 2018), sink to defects (Inoue, Ogura et al. 1977,

Choo and Lee 1982, Hong and Lee 1983, Languillaume, Kapelski et al. 1997), crack nucleation site (Cooke and Beevers 1974, Taylor, Warin et al. 1977) and propagation path (Taylor, Warin et al. 1977, Bae, Nam et al. 1996). Nevertheless, most studies (Embury and Fisher 1966, Langford 1970, Hyzak and Bernstein 1976, Langford 1977, Sevillano 1991, Embury and Hirth 1994, Janecek, Louchet et al. 2000, Zhang, Godfrey et al. 2011, Karkina, Karkin et al. 2015, Kapp, Hohenwarter et al. 2016) consider the FCI as a mere barrier for the lattice dislocations to explain the hardening behavior of pearlitic steel based on dislocation pile-up mechanism.

Pearlite colony is formed by diffusive eutectoid reaction from parent austenite ($\gamma$-Fe, face centered-cubic structure) phase into resultant ferrite and cementite phase (Soffa and Laughlin 2014). During the phase transformation, ferrite and cementite form specific crystallographic ORs and habit planes to minimize the energy barrier to phase transformation (Zhang and Kelly 1998, Zhang and Kelly 2005, Zhang, Esling et al. 2007). For a few decades, Isaichev (IS), Bagariatsky (BA), and Pitsch-Petch (PP) ORs had been known as three major ORs at the FCI (Isaichev 1947, Bagaryatsky 1950, Petch 1953, Pitsch 1962, Andrews 1963, Zhou and Shiflet 1991, Zhou and Shiflet 1992). However, recent high resolution experimental studies (Zhang, Esling et al. 2007) found that three major ORs are IS, Near BA, and Near PP ORs, and our recent computational study (Kim, Kang et al. 2016) also confirmed that Near BA and Near PP ORs has lower interfacial energies than BA or PP ORs (see Supplementary Figure S1-4). The Near BA and Near PP correspond to near Bag and P-P I-1 ORs denoted by Zhang et al. (Zhang, Esling et al. 2007), respectively. Our previous study (Kim, Kang et al. 2016) characterized the misfit dislocations at the FCI for the five ORs (IS, BA, PP, Near BA, and Near PP), and found that very different misfit dislocation networks are formed at the FCI in relatively similar ORs (such as PP and Near PP, BA and Near BA). In addition, there have been many studies (Bowden and Kelly 1967, Morgan and Ralph 1968, Shackleton and Kelly 1969, Dippenaar and Honeycombe 1973, Schastlivtsev and Yakovleva 1974, Sukhomlin 1976,

Mangan and Shiflet 1999, Zhang, Wang et al. 2012, Zhang, Zhang et al. 2012) to understand the occurrence of specific ORs in pearlitic steel. It has been reported that the occurrence of specific OR in pearlitic steel can be controlled by carbon contents of the system (Dippenaar and Honeycombe 1973), heat treatment condition (Mangan and Shiflet 1999) and applying magnetic field during phase transformation (Zhang, Wang et al. 2012, Zhang, Zhang et al. 2012).

A few experimental studies (Inoue, Ogura et al. 1977, Languillaume, Kapelski et al. 1997) reported that dislocation density is high near the FCI while it is low within the ferrite layer, and suggested that the FCI may act as a strong sink of lattice dislocations existing at the vicinity of the FCI. This lattice dislocation trapping mechanism in fine lamellar structure is known to be caused by the interfacial shear by the stress field of the lattice dislocation approaching to the interface, and thus, the dislocation trapping ability depends on the interfacial strength. Once the lattice dislocation is absorbed into the interface by shearing the interface, the lattice dislocation cannot escape the interface easily because the core structure of lattice dislocation spreads along the interface. This gives rise to extremely high strength of multi-layered metallic composite with a few nanometers of layer thickness (Hoagland, Hirth et al. 2006). Analogously, the strength of pearlitic steel with ultra-fine lamellar structure is attributed to the lattice dislocation trapping mechanism of the FCI. Moreover, we expect that the FCI structure for each OR will show different lattice dislocation trapping ability, because the characteristics of misfit dislocation for each OR is different (Kim, Kang et al. 2016) as well as the interfacial strength for each OR is different as discussed in the later part of the present paper. This suggests that FCI structure-property relationships can be varied depending on the crystallographic OR.

It was reported that pearlitic stel with a fine lamellar structure of a few tens to hundreds of nanometers can be formed by controlling the experimental conditions during the diffusive eutectoid reaction (Jaramillo, Babu et al. 2005, Wu and Bhadeshia 2012). Such nanostructured pearlites may have different mechanical strengths if different ORs dominate at the FCI, although their lamellar spacings are similar. In other words, if we can fabricate pearlitic steels with a fine lamellar structure with a desired OR, we can tune the mechanical properties of them based on the structure-property relationship of the FCI for each OR.

In order to reveal the structure-property relationship of the FCI for each OR, we investigated the effect of misfit dislocation at the FCI in different ORs on the interfacial strength under in-plane shear deformation, by employing atomistic simulations. We modeled the lamellar structures of pearlite colony as ferrite/cementite bilayer for IS, Near BA and Near PP OR, and characterized the misfit dislocations at the FCI by combining the xAIFB and disregistry analysis. We conducted in-plane shear deformation along twelve different directions on the FCI to observe the in-plane shear response of the FCI for each OR. The simulation results revealed that IS and Near BA ORs show dislocation-mediated plasticity except two directions, while Near PP OR shows mode II (in-plane shear) fracture at the FCI along all directions. Our results showed that the in-plane shear behavior of the FCI is governed primarily by magnitude of Burgers vector and core-width of misfit dislocation.

## 2. Methods
### 2.1. Modeling of the initial ferrite/cementite interface

In order to study the in-plane shear behavior of the FCI in pearlitic steel, it is necessary to model the initial FCI. We idealized the lamellar structure of pearlite colony as a flat ferrite/cementite bilayer. In the atomic simulation, we used modified embedded-atom method (MEAM) potential developed by Liyanage (Liyanage, Kim et al. 2014) to consider the directional interatomic interaction among Fe and C atoms. The lattice parameter of body-centered cubic (BCC) ferrite is $a_f$ = 2.851 Å and the lattice parameters of orthorhombic cementite are $a_c$ = 4.470 Å, $b_c$ = 5.088 Å and $c_c$ = 6.670 Å. To generate the initial configuration of ferrite/cementite bilayer, we determined the size of the perfect ferrite and cementite blocks to have minimal misfit strain for each OR. The detailed information of each OR and its habit plane are listed in Table 1. For the sake of simplicity, we intentionally put the $y$-axis perpendicular to the habit plane. To assure the periodic boundary conditions in $x$ and $z$-directions, we allowed small misorientation within 0.5 degree at most when necessary. Then, we iteratively applied biaxial strain to perfect ferrite and cementite blocks separately until satisfying both geometric compatibility and mechanical equilibrium condition. Finally, we assembled the strained ferrite and cementite blocks into the un-relaxed ferrite/cementite bilayer. When we assemble the strained ferrite and cementite blocks, we considered all the shuffle planes of cementite phase for each OR. To capture the initial state of the FCI, we carried out the molecular statics (MS) simulation using the conjugate gradient method to find energy minimum state of the ferrite/cementite bilayer. In order to find global minimum state, we performed the simulated annealing (SA) from 800 K to 10 K for 100 ps using Nosé-Hoover isobaric-isothermal (NPT) ensemble. After the SA procedure, we carried out MS simulation again to find the energy minimum state at 0 K. Finally, we can obtain the initial (relaxed) ferrite/cementite bilayer for each OR. The interface structure of un-relaxed and relaxed structure of the FCI for Near PP OR is described in Figure 1-a and b. In addition, we computed interface energies to choose the most stable interface structure among all shuffle planes of cementite (see Supplementary Figure S1-3). For BA and IS OR, there are several atomistic simulation studies on the most stable interface structure (Zhang, Hickel et al. 2015, Guziewski, Coleman et al. 2016, Zhou, Zheng et al. 2017) . The interface structures for BA and IS OR in the references match well with our simulation results (see Supplementary Figure S4 and S6).

We note that the residual stress in ferrite and cementite block induced by the biaxial strain can affect the interface energy and shear strength of the FCI calculated here, which would induce an intrinsic size effect. Indeed, because of the high computational cost of performing interfacial shear simulation, we used smaller simulation cells compared to those of our previous study (Kim, Kang et al. 2016). However, the interface energy increases at most by 4% upon the size reduction, and the characteristics of the misfit dislocation network are almost identical. Because the interfacial strength depends on the Burgers vector and core-width of the misfit dislocations, we expect that the qualitative findings in the present study such as different failure modes for different ORs do not change over the increase of simulation cell. The detailed information such as size, misfit strain and the lowest interface energy can be found in the Table 2 and schematic diagram of the initial ferrite/cementite bilayer for each OR is presented in Figure 2-a.

### 2.2. In-plane shear deformation of the ferrite/cementite interface

In order to investigate the in-plane shear behavior of the FCI, we applied the simple shear deformation under force control at 300 K. Before applying the shear deformation, we performed equilibration procedure to the initial ferrite/cementite bilayer using NPT ensemble at 300 K for 200 ps. To impose the simple shear condition to the bilayer model, we assigned the in-plane net force **F** (only with *x* and *z*-components) on the top and bottom slabs with same magnitude but in opposite directions as described in Figure 2. Total 12 different in-plane shear deformations with different direction angle $\theta$ were performed to capture the anisotropic shear response of the FCI for each OR. Where direction angle $\theta$ represents the angle between positive *x*-axis and direction of net force assigned on top slab around negative *y*-axis as depicted in Figure 2-a. The magnitude of the in-plane force applied to the top and bottom slabs was determined by desired shear stress at a given loading step. The shear stress is monotonically increased by increment of shear stress, Δτ, at each loading step. For each loading step, we performed NPT ensemble to eliminate all stress components except for $\tau_{xy}$ and $\tau_{yz}$ with given in-plane net force at 300 K for 200 ps.

### 2.3. Core-width of misfit dislocation at the ferrite/cementite interface

In order to understand the anisotropic shear response of the FCI, we analyzed the characteristics of misfit dislocation at the FCI. To characterize the misfit dislocation of the FCI in terms of Burgers vector (**b** = [$b_e$, $b_s$]), line orientation (ξ) and line spacing (*d*), it is essential to determine a reference lattice structure to both ferrite and cementite structures. We perform the xAIFB analysis (Kim, Kang et al. 2016) on the initial FCI structures to find a reference lattice structure that make the net Burgers vector from the Frank-Bilby equation and Knowle's equation equal to each other, and thus to determine the unique Burgers vector. In addition, Peierls-Nabarro (P-N) model (Peierls 1940, Nabarro 1947) indicates that not only the magnitude of Burgers vector ($b=|\mathbf{b}|$) but also the core-width ($2w$) play an important role to determine the Peierls stress which represents the critical shear stress to activate the glide of lattice dislocation in bulk. Analogously, if the in-plane motion of the FCI misfit dislocation is considered as glide of a lattice dislocation, how wide the core spreads must have a significant effect on the in-plane shear strength of the FCI. So, we performed the disregistry analysis (Kang, Wang et al. 2012, Wang, Zhang et al. 2013, Wang, Zhang et al. 2014, Kim, Kang et al. 2016) to measure the core-width of misfit dislocation at the FCI. To this end, we used disregistry $r_{ij}^{rc}$, the relative displacement from the reference structure to relaxed one. Each dislocation causes a stepwise change in disregistry $r_{ij}^{rc}$. The magnitude of Burgers vector and core-width determine the height of the step and transition from core to adjacent core, respectively. In order to measure the core-width of dislocation, we fitted the disregistry $r_{ij}^{rc}$ to the following stepwise function:

$$r_\lambda^{rc}(q) = \sum_{\beta=1}^{N_{dis}} \left\{ \frac{b_\lambda^\beta}{\pi} \sum_{\alpha=1}^{n} \text{atan}\left(\frac{q - \alpha p_\lambda - u_\lambda}{w_\lambda^\beta}\right) \right\} + v_\lambda \qquad (1)$$

where $r_\lambda^{rc}$ indicates the non-uniform displacement from reference lattice structure to initial (relaxed) ferrite/cementite bilayer at the interface projected onto the *s* or *t*-axis for $\lambda$ component. **s** and **t** are probe vectors perpendicular and paralleled to dislocation line ($\xi$) on the interface, respectively (see Figure 1-c and d). **s** can be determined as $\mathbf{s} = \mathbf{t} \times \mathbf{n}$, where **n** is paralleled to *y*-axis. $\lambda$ represents the component of each misfit dislocation and the subscript e and s represent the edge and screw components of each misfit dislocation, respectively. *q* represents position of given data point projected onto the probe axis *s* or $q = \mathbf{x} \cdot \mathbf{s}$. $b_\lambda^\beta$ and $w_\lambda^\beta$ represent $\lambda$ component of Burgers vector and half-width of dislocation core for $\beta^{th}$ misfit dislocation, respectively. $p_\lambda$, $u_\lambda$ and $v_\lambda$ represent the period, shift in *s*- and $r_\lambda^{rc}$-axis, respectively. $N_{dis}$ and *n* represent the number of misfit dislocations and periodic function, respectively. The schematic diagram of disregistry analysis using Equation 1 is given in Figure 3-a. There are five unknowns ($b_\lambda^\beta$, $w_\lambda^\beta$, $p_\lambda$, $u_\lambda$ and $v_\lambda$) to define the shape of disregistry. In order to determine the shape of disregistry, we performed the optimization procedure to minimize the cost function, *f*, defined as

$$f(b_\lambda^\beta, w_\lambda^\beta, p_\lambda, u_\lambda, v_\lambda) = \sum_{\alpha=1}^{N_{atom}} \left\{ r_{ij,\lambda}^{rc,\,atom}(q_\alpha) - r_\lambda^{rc}(q_\alpha) \right\}^2 \qquad (2)$$

where $N_{atom}$ is the number of data points of disregistry obtained from atomistic simulation. $r_{ij,\lambda}^{rc,\,atom}$ indicates the $\lambda$ component of disregistry projected onto the local axis (*s*- or *t*-axis) for $\alpha^{th}$ disregistry point. $q_\alpha$ indicates the position projected onto *s*-axis for $\alpha^{th}$ disregistry point. To find the five unknown variables with reasonable tolerance, we used *fmincon* function with trust-region-reflective algorithm implemented in the MATLAB R2016b.

## 2.4. Analysis of shear deformation of the ferrite/cementite interface

Due to different shear moduli of ferrite and cementite, different shear strains are developed for each phase for a shear stress imposed and it accompanies ambiguity in determining the shear strain at the interface region during shear deformation. Hence, it is necessary to define a new measure of shear strain at the interface to analyze the shear behavior of the FCI. We assume that ferrite and cementite experience different uniform shear strains in each phase and there is no displacement in $y$-direction and that the in-plane displacement ($x$ and $z$-component) at a given point linearly depends on the $y$-coordinate (see Figure 2-c and d). Based on these assumptions, we defined average in-plane shear strains, $\gamma_{xy}$ and $\gamma_{yz}$, as shown below

$$u^\kappa = m_x^\kappa \cdot \left(y^\kappa - y_p^\kappa\right) + u_p^\kappa \text{ and } w^\kappa = m_z^\kappa \cdot \left(y^\kappa - y_p^\kappa\right) + w_p^\kappa, \text{ where } \kappa = \text{f or c}$$

$$\gamma_{xy} = \left(u_{int}^c - u_{int}^f\right)/h, \quad \gamma_{yz} = \left(w_{int}^c - w_{int}^f\right)/h, \quad h = y_{int}^c - y_{int}^f \tag{3}$$

where $\kappa$ represents the phase at a given $y$-coordinate and the superscripts f and c indicate the ferrite and cementite phase, respectively. $u^\kappa$ and $w^\kappa$ represent the average in-plane displacement along the $x$ and $z$-axis at $y = y^\kappa$ for $\kappa$ phase, respectively. $u_p^\kappa$ and $w_p^\kappa$ are the in-plane displacement in $x$ and $z$-directions, respectively, at the reference point at $y = y_p^\kappa$ for $\kappa$ phase. $m_x^\kappa$ and $m_z^\kappa$ are the slope of the in-plane displacement in $x$ and $z$-directions for $\kappa$ phase, respectively. We applied linear regression using multiple reference points to compute the $m_x^\kappa$ and $m_z^\kappa$ value. $\gamma_{xy}$ and $\gamma_{yz}$ are the average shear strain at the interface region and $h$ represents the average interface spacing between topmost ferrite layer and bottommost cementite layer in $y$-direction (see Figure 2). The average interface spacing $h$ were 1.30, 1.33 and 1.24 Å for IS, Near BA and Near PP OR, respectively.

In order to analyze the in-plane shear deformation at the FCI, we computed the disregistry $r_{ij}^r$, which represents the difference between displacement of $i^{th}$ atom in ferrite (or cementite) and displacement of $j^{th}$ atom in cementite (or ferrite) for each loading step. The displacement of $i^{th}$ atom in κ phase represents the displacement from the initial atomic position to the deformed atomic position at the interface region. κ represents either ferrite or cementite phase.

## 3. Results

### 3.1. The characteristics of misfit dislocations for three orientation relationships

Having obtained the atomic structure of the FCI, we performed the xAIFB and disregistry analysis to determine the characteristics of in-plane misfit dislocations such as λ components of Burgers vector $b_\lambda^\beta$ and half-width of dislocation core $w_\lambda^\beta$, line orientation vector $\xi^\beta$ and line spacing $d^\beta$ for $\beta^{th}$ misfit dislocation. For IS OR, we set the crystallographic orientation as $\mathbf{x} = [010]_c \| [\bar{1}11]_f$, $\mathbf{y} = (\bar{1}01)_c \| (\bar{1}\bar{2}1)_f$ and $\mathbf{z} = \mathbf{x} \times \mathbf{y}$ and generated the perfect ferrite and cementite blocks with given orientation, separately. The subscript f and c represent the ferrite and cementite phases, respectively. For IS OR, the dimensions of the initial ferrite/cementite bilayer were $L_x$ = 167.30 Å, $L_y^c$ = 81.69 Å, $L_y^f$ = 83.80 Å and $L_z$ = 32.12 Å. $L_x$ and $L_z$ represent the length of the simulation box for the ferrite/cementite bilayer in $x$ and $z$-directions, respectively. $L_y^c$ and $L_y^f$ represent the length of the cementite and ferrite block in $y$-direction, respectively. The misfit strains between ferrite and cementite block for IS OR were $\varepsilon_{xx}$ = 9.08×10$^{-6}$ and $\varepsilon_{zz}$ = 2.17×10$^{-3}$ in $x$- and $z$-directions, respectively. Habit plane of cementite phase has 8 different shuffle planes for IS OR. Among the 8 different shuffle planes, the lowest interface energy of the FCI for IS OR was $\gamma^{cf}$ = 501.9 mJ/m². In the in-plane shear deformation, we used the FCI model which has the lowest interface energy (see supplementary Figure S1). In addition, recent experimental work (Zhou, Zheng et al. 2017) computed most stable interface structure using density functional theory and directly observed atomic configuration of the FCI for IS OR. From the reference, we found that they only considered five shuffle planes out of eight shuffle planes of cementite for IS OR (see supplementary Figure S5). Even though they missed several shuffle planes of cementite, the reported atomic configuration at the interface is well matched with atomic configuration of the FCI model for IS OR computed by MEAM potential (Liyanage, Kim et al. 2014) (see supplementary Figure S6). In order to show the local coherency and geometry of misfit dislocations at the FCI, we visualized the atomic potential energy maps for the interface atoms and presented the corresponding geometries of idealized misfit dislocation in Figure 4. As shown in Figure 4-a, a single array of edge dislocation was developed in $z$-direction on the FCI because of the atomic mismatch along the $x$-axis. The xAIFB and disregistry analysis reveal the characteristics of misfit dislocations. The edge component of Burgers vector was $b_e^1 = 2.53$ Å, and half-width of dislocation core was $w_e^1 = 13.21$ Å as plotted in Figure 3-b. The line orientation was $\xi^1 = 90°$ (paralleled to $z$-axis), and line spacing was $d^1 = 83.65$ Å (see Figure 4-a).

For Near BA OR, we set the crystallographic orientation as $\mathbf{x} = [010]_c \| [\bar{1}11]_f$, $\mathbf{y} = (001)_c \| (\bar{5}94)_f$ and $\mathbf{z} = \mathbf{x} \times \mathbf{y}$. The dimensions of the initial ferrite/cementite bilayer for Near BA OR were $L_x = 167.18$ Å, $L_y^c = 93.38$ Å, $L_y^f = 94.47$ Å and $L_z = 54.22$ Å. The misfit strains between ferrite and cementite blocks for Near BA OR were $\varepsilon_{xx} = 9.08 \times 10^{-6}$ and $\varepsilon_{zz} = 8.38 \times 10^{-3}$ in $x$- and $z$-directions, respectively. Cementite block for Near BA OR has 6 different shuffle planes. Among these, the lowest interface energy of the FCI for Near BA OR was $\gamma^{cf} = 539.1$ mJ/m² (see supplementary Figure S2). As shown in Figure 4-b, dislocation network in brick and mortar pattern was developed on the FCI with two edge dislocations. The edge components of Burgers vector were $b_e^1$ = 2.21 Å and $b_e^2$ = 2.53 Å and the half-widths of dislocation core were $w_e^1$ = 2.12 Å and $w_e^2$ = 6.67 Å for each misfit dislocation, respectively (see Figure 3-c and d). The line orientations were $\xi^1$ = 0° (paralleled to x-axis) and $\xi^2$ = 90° (paralleled to z-axis), and the line spacings were $d^1$ = 18.07 Å and $d^2$ = 83.59 Å, respectively (see Figure 4-b).

For Near PP OR, we put the crystallographic orientation as $\mathbf{x}$ = $[010]_c \| [\bar{1}13]_f$, $\mathbf{y}$ = $(001)_c \| (24\ 9\ 5)_f$ and $\mathbf{z}$ = $\mathbf{x} \times \mathbf{y}$. The dimension of the initial ferrite/cementite bilayer for Near PP OR were $L_x$ = 65.77 Å, $L_y^c$ = 73.37 Å, $L_y^f$ = 74.45 Å and $L_z$ = 44.99 Å. The misfit strains between ferrite and cementite block with Near PP OR were $\varepsilon_{xx}$ = 3.66×10$^{-4}$ and $\varepsilon_{zz}$ = 2.24×10$^{-3}$ in x- and z-directions, respectively. Cementite phase for Near PP OR has 6 different shuffle planes. Among these, the lowest interface energy of the FCI for Near PP OR was $\gamma^{cf}$ = 575.9 mJ/m² (see supplementary Figure S3). As shown in Figure 4-c, the dislocation network composed of two straight dislocation was formed at the FCI. As shown in Figure 3-e and f, the edge components of Burgers vector were $b_e^1$ = 1.06 Å and $b_e^2$ = 3.26 Å and the half-widths of dislocation core were $w_e^1$ = 1.36 Å and $w_e^2$ = 4.04 Å for each dislocation line, respectively. The screw components of Burgers vector were $b_s^1$ = 2.18 Å and $b_s^2$ = 3.72 Å and the half-widths of dislocation core were $w_s^1$ = 2.00 Å and $w_s^2$ = 4.37 Å for each dislocation line, respectively. The line orientations were $\xi^1$ = 0° (paralleled to x-axis) and $\xi^2$ = 41.3°, and the line spacings were $d^1$ = 22.35 Å and $d^2$ = 43.41 Å, respectively (see Figure 4-c). The detailed information on the characteristics of misfit dislocation for three FCIs are summarized in Table 3.

### 3.2. In-plane shear response of the ferrite/cementite interface

Figure 5 presents the in-plane shear resistance maps for twelve different shear directions and stress-strain curves for $\theta = 0°$ for three ORs. The distance from each data point to origin represents the magnitude of the in-plane shear resistance as depicted in Figure 5. From Figure 5-a, we noticed that in-plane shear resistance map shows significant anisotropy for three ORs. In-plane shear resistance map for IS OR has largest anisotropy while Near PP has the smallest. For IS OR, the minimum shear resistance among twelve directions is around 100 MPa and the maximum shear resistance is around 9.18 GPa. The maximum in-plane shear resistance is out of the range of the Figure 5-a. Figure 6 contains the potential energy map to identify the initial geometry of misfit dislocation and disregistry $r_{ij}^r$ maps for $\theta = 0°$ to inspect the deformation behavior of the FCI. At low stress level for $\theta = 0°$ ($\tau_{xy} \approx 50$ MPa), the disregistry is relatively uniformly distributed along the misfit dislocation core region. At high stress level for $\theta = 0°$ ($\tau_{xy} \approx 100$ MPa), the disregistry $r_{ij}^r$ is more concentrated near the core-region of misfit dislocation (see Figure 5-b and 6-a). In addition, Figure 7 shows a series of snapshot at the FCI during in-plane shear deformation for $\theta = 0°$. As shown in Figure 7-a, we observed the glide of the misfit dislocation on the interface plane at the FCI for IS (see supplementary Movie 1). We also compute relative displacements of interface atoms in x-direction under constant shear stress for $\theta = 0°$ ($\tau_{xy} \approx 110$ MPa for IS OR, $\tau_{xy} \approx 570$ MPa for Near BA OR and $\tau_{xy} \approx 1870$ MPa for Near PP OR) to confirm if the deformation is produced by misfit dislocation motion or interfacial fracture (see Figure 8). For easy comparison, we set the initial relative displacement as zero right after desired shear stress is applied. Relative displacement for IS OR linearly increased with time because of the glide of misfit dislocation, which corresponds to the mobility law of dislocation in bulk metal under constant shear stress. The FCI for IS OR plastically deforms by misfit dislocation glide on the interface except for $\theta = 90°$ and $270°$. Unlike other shear directions, the FCI for $\theta = 90°$ and $270°$ shows mode II (in-plane shear) fracture at the interface.

For Near BA and Near PP ORs, the minimum shear resistance among twelve directions is around 380 MPa and around 1300 MPa, respectively, and the maximum shear resistance is around 4.66 GPa and 6.51 GPa, respectively (see Figure 5-a and b). At low stress level for $\theta = 0°$ ($\tau_{xy} \approx 190$ MPa for Near BA OR and $\tau_{xy} \approx 560$ MPa for Near PP OR), we observe that the disregistry is locally initiate at the junction of two different misfit dislocation lines (see Figure 6-b and c). At high stress level for $\theta = 0°$ ($\tau_{xy} \approx 380$ MPa for Near BA OR and $\tau_{xy} \approx 1700$ MPa for Near PP OR), the disregistry $r_{ij}^r$ spreads out to adjacent core-regions of misfit dislocations (see Figure 6-b and c). As shown in Figure 7-b and c, the FCI for Near BA OR shows glide of the misfit dislocation on the interface while the FCI for Near PP OR shows mode II fracture for $\theta = 0°$ (see supplementary Movie 2 and 3). Figure 8 also shows that relative displacement at the interface atoms of Near BA OR monotonically increased with time because of glide of misfit dislocation on the FCI while relative displacement of Near PP OR shows a sudden increase due to local fracture at the FCI. Moreover, similar to in-plane shear behavior of the FCI for IS OR, the FCI for Near BA OR shows glide of misfit dislocation on the interface except for $\theta = 90°$ and $270°$ where mode II fracture is observed. Unlike the others, the FCI for Near PP OR shows mode II fracture for all in-plane shear directions.

## 4. Discussion

### 4.1. In-plane shear resistance of Isaichev orientation relationship (IS OR)

For IS OR, in-plane shear resistance of the FCI shows highest anisotropy among the three different ORs as plotted in Figure 5-a. The in-plane shear resistance to the $\tau_{xy}$ component is nearly constant ($\tau_{xy} \approx 100$ MPa) and $\tau_{yz}$ component of in-plane shear resistance shows its highest value of $\tau_{yz} \approx 9.18$ GPa when $\theta = 270°$. The extremely anisotropic in-plane shear response of the ferrite/cementite bilayer for IS OR can be explained by the magnitude of Burgers vector and core-width of the misfit dislocation. The FCI contains an array of edge dislocations as depicted in Figure 4-a. The Peach-Koehler (P-K) force **f** acting on the misfit dislocation under in-plane shear is given as $\mathbf{f} = [b_e^1 \tau_{xy}, 0, 0]$ where $b_e^1$ and $\tau_{xy}$ represent the edge component of Burgers vector and $xy$-component of externally applied shear stress. Hence, when applied shear stress has $xy$-component, the shear strength is estimated from the condition of $\tau_{xy} > \tau_p$ where $\tau_p$ represents the Peierls-type critical stress. Although the concept of Peierls stress is originally applied to an infinitely straight single lattice dislocation gliding on the slip plane in bulk, in this study, we adopt it to estimate the required stress to activate the motion of the misfit dislocation on the FCI. As the direction angle of in-plane shear stress is $\theta$ to the $x$-axis (i.e. $\tau_{xy} = \tau \cos\theta$ where $\tau$ represents the magnitude of applied in-plane shear stress), interfacial strength is given as $\tau_p / \cos\theta$, which well describes the anisotropic interfacial strength map of IS OR in Figure 5-a. According to the P-N model (Peierls 1940, Nabarro 1947, Lubarda and Markenscoff 2006, Lubarda and Markenscoff 2006), the Peierls stress is expressed as $\tau_p = 2\mu/ab\ [b_e^2/(1-\nu) + b_s^2] \exp(-4\pi w_e/b)$ where $\mu$ and $\nu$ represent effective shear modulus and Poisson's ratio, respectively. $w_e$, $a$ and $b$ represent edge component of half-width of dislocation core, interatomic distance within the slip plane in direction perpendicular to the dislocation line and magnitude of Burgers vector, respectively. $b_e$ and $b_s$ represent the edge and screw components of Burgers vector, respectively. P-N model indicates that the Peierls stress $\tau_p$ becomes smaller when $w_e/b$ becomes larger. In other words, the wide core region with small magnitude of Burgers vector gives rise to a small in-plane shear resistance of the FCI. In the P-N model, the variable $w_e/b$ of exponential function has a dominant effect in Peierls stress compared to other factors. So, we used the ratio of half-width of dislocation core to magnitude of Burgers vector, $w_e/b$, to explain the trend of in-plane shear resistance of the FCI in different ORs. The misfit dislocation developed at the FCI in IS OR has the largest ratio of dislocation core half-width to Burgers vector magnitude ($w_e^1/b^1 = 5.22$), which explains the smallest in-plane shear resistance of IS OR among three different ORs. On the other hand, the maximum in-plane shear resistance among the three different FCIs also can be found in the FCI for IS OR. The IS OR shows 9.18 GPa of maximum in-plane shear resistance when applied shear stress has only *yz*-component. The large in-plane shear resistance along yz-direction is originated from the absence of P-K force along the direction. Hence, the interfacial strength is determined by the coherency of atoms at the FCI for the case. We found that the disregistry is concentrated at the core-region of misfit dislocation just before the fracture. It implies that the fracture is initiated at the core-region due to the incoherency near the core-region.

We note that a recent experimental study (Zhou, Zheng et al. 2017) predicts two sets of misfit dislocation arrays in both x and y directions with 25 and 50 nm line spacing, which is different from our simulation results predicting a misfit dislocation array along one direction. The discrepancy may originate from the small errors in the predicted lattice constants from the MEAM potential. In addition, the lattice constants at eutectoid temperature are different from those at room temperature while we construct the simulation cell without considering the temperature effect. Although one can resolve the discrepancy related to the lattice constants, a larger simulation involving 50 nm spacing misfit dislocations is too computationally expensive. Since the absence of misfit dislocation along one direction may have a discernible effect on the results of the Isaichev OR, depending on the characteristic of another misfit dislocation array, the in-plane shear strength map in Figure 5-a may change if it is obtained with a larger simulation.

4.2. In-plane shear resistance of Near Bagaryatsky and Near Pitsch-Petch orientation

relationships

For Near BA and Near PP ORs, in-plane shear resistance of the FCI also shows significant anisotropy as depicted in Figure 5-a. Unlike the FCI in IS OR, the FCIs for Near BA and Near PP ORs contain two types of misfit dislocations with different characteristics. Figure 4 shows that the multiple junctions of misfit dislocations are developed at the FCI from the interaction of two different misfit dislocations. In Figure 6-b and c, the disregistry for $\theta = 0°$ is initiated at the junction of the misfit dislocation at relatively low shear stress level ($\tau_{xy} \approx$ 190 MPa for Near BA OR and $\tau_{xy} \approx$ 560 MPa for Near PP OR). The localized initial shear deformation can be understood from the potential energy map. The potential energy maps for Near BA and Near PP ORs show higher potential energy near the junction region (Figure 6-b and c), which implies that relatively weak atomic bonding is present at the junction formed by the reaction of two misfit dislocations.

For Near BA OR, the in-plane shear resistance map shows that the in-plane shear resistance to the $\tau_{xy}$ component is nearly constant ($\tau_{xy} \approx 350$ MPa) while the resistance to the pure $\tau_{yz}$ stress has the highest value ($\tau_{yz} \approx 4.66$ GPa) as plotted in Figure 5-a. From the atomistic simulation results, we observed that the FCI for Near BA OR shows glide of misfit dislocation on the interface except for $\theta = 90°$ and $270°$ where mode II fracture is observed. According to the P-K equation, forces exerted on the first and second misfit dislocations are $\mathbf{f}_1 = [0, 0, b_e^1 \tau_{yz}]$ and $\mathbf{f}_2 = [b_e^2 \tau_{xy}, 0, 0]$, respectively. $\tau_{yz}$ and $\tau_{xy}$ represent $yz$ and $xy$-components of applied in-plane shear stress, and $b_e^1$ and $b_e^2$ represent the Burgers vectors for two edge dislocations. If both misfit dislocations are glissile, the shear deformation will be initiated when if either $\tau_{yz} > \tau_p^1$ or $\tau_{xy} > \tau_p^2$ is satisfied. $\tau_p^1$ and $\tau_p^2$ represent the Peierls stresses of first and second dislocations depicted in Figure 4. The shear strength can be predicted as a function of the angle $\theta$, as $\min\left(\tau_\mathrm{p}^1/\sin\theta, \tau_\mathrm{p}^2/\cos\theta\right)$. The ratio of the half-width of dislocation core to magnitude of Burgers vector for each misfit dislocation ($w_\mathrm{e}^1/b^1 = 0.96$ and $w_\mathrm{e}^2/b^2 = 2.64$) implies that the second misfit dislocation has significantly lower Peierls stress than the first does. In other words, the critical shear stress to activate the motion of second misfit dislocation is significantly lower than that of the other, i.e. $\tau_\mathrm{p}^2 \ll \tau_\mathrm{p}^1$. Hence, the interfacial shear strength can be approximated as $\tau_\mathrm{p}^2/\cos\theta$ unless $\theta$ is close to 90° or 270°, which explains nearly constant $\tau_{xy}$ component of the in-plane shear resistance in Figure 5-a. Mode II fracture at $\theta = $ 90° and 270° is originated from large Peierls stress of the first misfit dislocations. We suspect that fracture occurs because the critical shear stress to promote the motion of first misfit dislocation is higher than the critical shear stress to initiate fracture at the FCI.

For Near PP OR, the in-plane shear strength is less anisotropic and significantly larger than IS or Near BA ORs, ranging from 1.30 to 6.51 GPa (Figure 5-a). The disregistry analysis reveals that the ratio of the half-width of dislocation core to magnitude of Burgers vector $w_\mathrm{e}/b$ are very small for both dislocations ($w_\mathrm{e}^1/b^1 = 0.56$ and $w_\mathrm{e}^2/b^2 = 0.82$). We suspect that, for both dislocations, the critical shear stress to activate the glide of misfit dislocation on the FCI is larger than the critical shear stress to initiate fracture at the FCI. Thus, the FCI for Near PP OR shows mode II fracture rather than dislocation-mediated plasticity for all in-plane shear directions.

From the simulation results, we conclude that the in-plane shear responses of the FCIs in IS and Near BA ORs are similar to each other in that shear deformation is mainly governed by dislocation motion. Since in-plane shear strength of the FCI for IS and Near BA OR is considerably small (a few hundred mega-pascals), the FCI may operate as a strong sink to the lattice dislocation. The high dislocation density near the FCI region (Inoue, Ogura et al. 1977, Languillaume, Kapelski et al. 1997) is likely attributed to the weak interfacial shear strength of the FCI in IS or Near BA ORs. It is interesting to note that the IS and Near BA OR have exactly same OR except different habit planes. Moreover, IS and Near BA OR are usually found in the same pearlite colony while Near PP OR is found in different pearlite colony, as reported in the literature (Zhang, Esling et al. 2007). It means that the FCIs in the pearlite colony in IS and Near BA OR has low in-plane shear resistance to externally applied shear stress, while the FCI in the pearlite colony in Near PP OR has high in-plane shear resistance to externally applied shear stress. Furthermore, the magnitude of the in-plane shear strength of the FCI for IS and Near BA OR is similar to the Peierls stress of the ferrite with lattice dislocation (a few hundred megapascals). Therefore, for IS and Near BA OR, in addition to glide of lattice dislocation in the ferrite phase, glide of misfit dislocation may contribute to the plastic response. For Near PP OR, since the in-plane shear resistance of the FCI for Near PP OR (a few gigapascals) is much higher than the Peierls stress of lattice dislocation in the ferrite lamellar, glide of lattice dislocation will govern the overall strain hardening behavior of the pearlitic steel with fine lamellar structure.

As mentioned earlier, the occurrence of specific OR in pearlitic steel can be controlled by carbon contents of the system, heat treatment condition, and application of the magnetic field during the phase transformation stage. If we can fabricate micro-pillar specimen from a colony of such-fabricated pearlites, we can perform compression tests of the specimen with specific FCI ORs to assess the present predictions, following an experimental procedure described in the literature (Kapp, Hohenwarter et al. 2016).

## 5. Conclusions

We study the in-plane shear response of the FCI for three ORs by using atomistic simulation methods combined with xAIFB and disregistry analyses. We find that the magnitude of Burgers vector and core-width of misfit dislocation govern the overall in-plane shear response of the FCI. We find that misfit dislocation is glissile when the $w_e/b$ ratio is large, but they become sessile when the ratio is small. The anisotropy of in-plane shear strength can be understood by different $w_e/b$ ratios of misfit dislocations along different directions. When dislocation is sessile (first and second dislocations of Near PP OR and first dislocation of Near BA OR) or P-K force is zero for specific direction ($\theta = 90°$ and $270°$ in IS OR), mode II fracture behavior is observed. For Near BA and Near PP OR where two sets of misfit dislocations intersect, we found that both plastic deformation and fracture initiated at the junction of misfit dislocations because of the formation of weak bonds at the junction. As a future work, we plan to study the lattice dislocation trapping at the FCI for each OR by investigating the local stress distribution near the FCI and stress field induced by lattice dislocation approaching to the FCI.

To summarize.

- The FCIs for IS, Near BA and Near PP ORs show different degrees of anisotropy in the in-plane shear resistance.

- The FCI for IS and Near BA OR shows dislocation-mediated plasticity for all in-plane shear directions except mode II fracture for $\theta = 90°$ and $270°$.

- The FCI for Near PP OR shows mode II fracture for all in-plane shear direction.

- Glide of misfit dislocation at the FCI for IS and Near BA OR can be explained by magnitude of Burgers vector and core-width of misfit dislocation.

- Mode II fracture for three ORs is initiated at the dislocation core-region because of the incoherency.

- Despite the presence of misfit dislocation, mode II fracture at the FCI can occur when

the required shear stress to activate the glide of misfit dislocation is higher than the required shear stress to initiate mode II fracture at the interface.

- The overall in-plane shear response of the FCI is governed by the magnitude of Burgers vector and core-width of misfit dislocation.

**Acknowledgments**

This research was supported by Basic Science Research Program (2016R1C1B2011979 and 2016R1C1B2016484) through the National Research Foundation of Korea (NRF) funded by the Ministry of Science, ICT & Future Planning.

**Tables**

Table 1. The geometry of the initial ferrite/cementite bilayer and the interface energy for each orientation relationship.

| Name    | $L_x$  | $L_y^c$ | $L_y^f$ | $L_z$ | $\varepsilon_{xx}$     | $\varepsilon_{zz}$     | $\gamma^{cf}$ [mJ/m$^2$] |
|---------|--------|---------|---------|-------|------------------------|------------------------|--------------------------|
| IS      | 167.30 | 81.69   | 83.80   | 32.12 | $9.08 \times 10^{-6}$  | $2.17 \times 10^{-3}$  | 501.9                    |
| Near BA | 167.18 | 93.38   | 94.47   | 54.22 | $9.08 \times 10^{-6}$  | $8.38 \times 10^{-3}$  | 539.1                    |
| Near PP | 65.77  | 73.37   | 74.45   | 44.99 | $3.66 \times 10^{-4}$  | $2.24 \times 10^{-3}$  | 575.9                    |

*Unit: Å (Angstrom)

Table 2. Detailed information regarding misfit dislocation for each orientation relationship.

| Name | Notation | IS | Near BA | Near PP |
|---|---|---|---|---|
| Magnitude of edge and screw component of Burgers vector | $[b_e^1, b_s^1]$ | [2.53, 0.00] | [2.21, 0.00] | [1.06, 2.18] |
| | $[b_e^2, b_s^2]$ | N/A | [2.53, 0.00] | [3.26, 3.72] |
| Magnitude of Burgers vector | $b^1$ | 2.53 | 2.21 | 2.42 |
| | $b^2$ | N/A | 2.53 | 4.95 |
| Half-width of dislocation core for edge component | $w_e^1$ | 13.21 | 2.12 | 1.36 |
| | $w_e^2$ | N/A | 6.67 | 4.04 |
| Half-width of dislocation core for screw component | $w_s^1$ | – | – | 2.00 |
| | $w_s^2$ | N/A | – | 4.37 |
| Half-width of dislocation core /magnitude of Burgers vector | $w_e^1/b^1$ | 5.22 | 0.96 | 0.56 |
| | $w_e^2/b^2$ | N/A | 2.64 | 0.82 |
| Line orientation | $\xi^1$ (deg.) | 90.00° | 0.00° | 0.00° |
| | $\xi^2$ (deg.) | N/A | 90.00° | 41.30° |
| Line spacing | $d^1$ | 83.65 | 18.07 | 22.35 |
| | $d^2$ | N/A | 83.59 | 43.41 |

*Unit: Å (Angstrom), line orientation represents the angle between positive $x$-axis and dislocation line around negative $y$-axis.

*superscript e and s represents the component of the edge and screw component for each dislocation line, respectively.

**Figures**

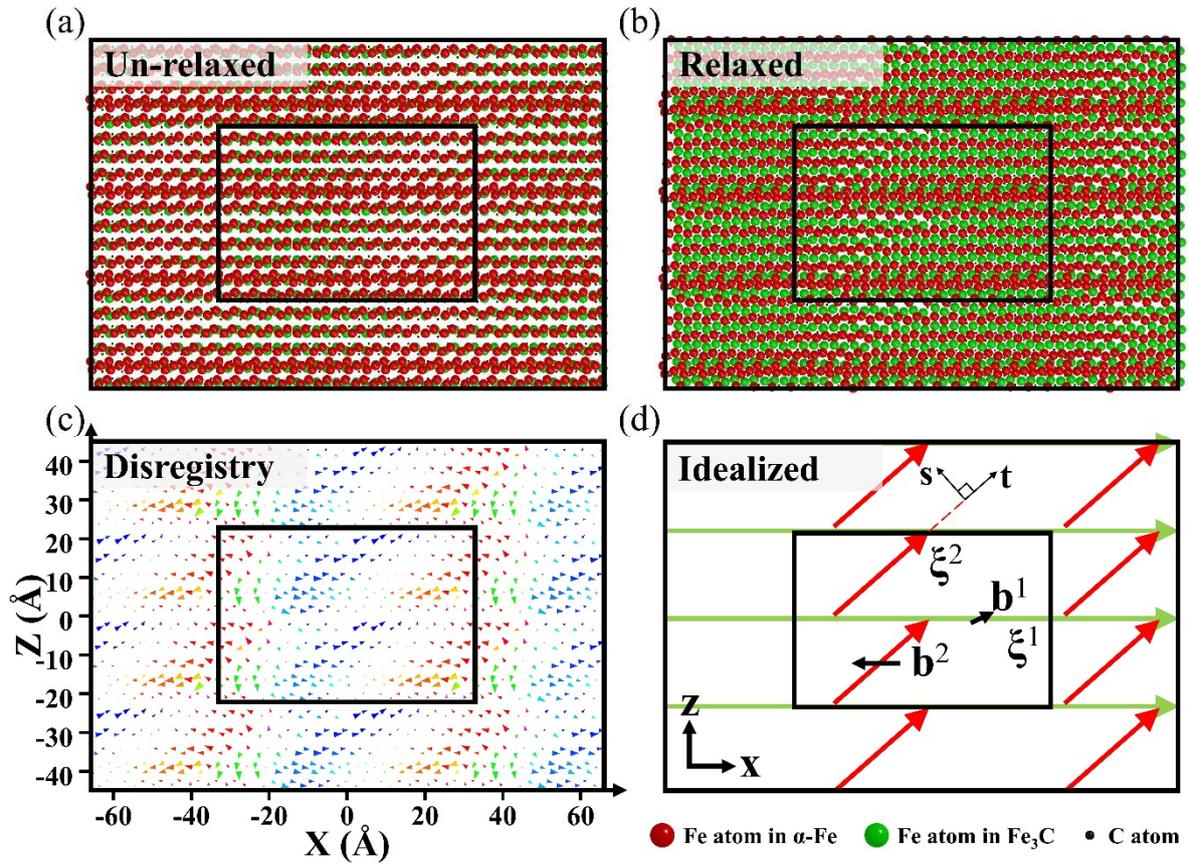

Figure 1. (a) Un-relaxed, (b) relaxed structure, (c) 2D disregistry map and (d) idealized misfit dislocations of the ferrite/cementite interface for Near Pitsch-Petch (Near PP) orientation relationship; Black rectangular box in each figure represents the simulation box.

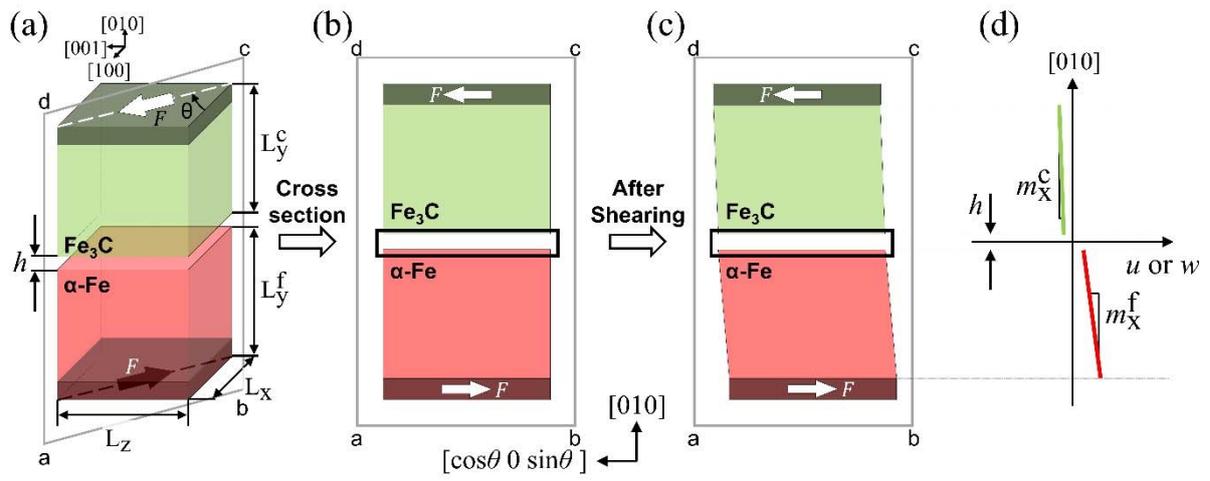

Figure 2. Schematic representation of in-plane shear deformation of the ferrite/cementite bilayer for each orientation relationship

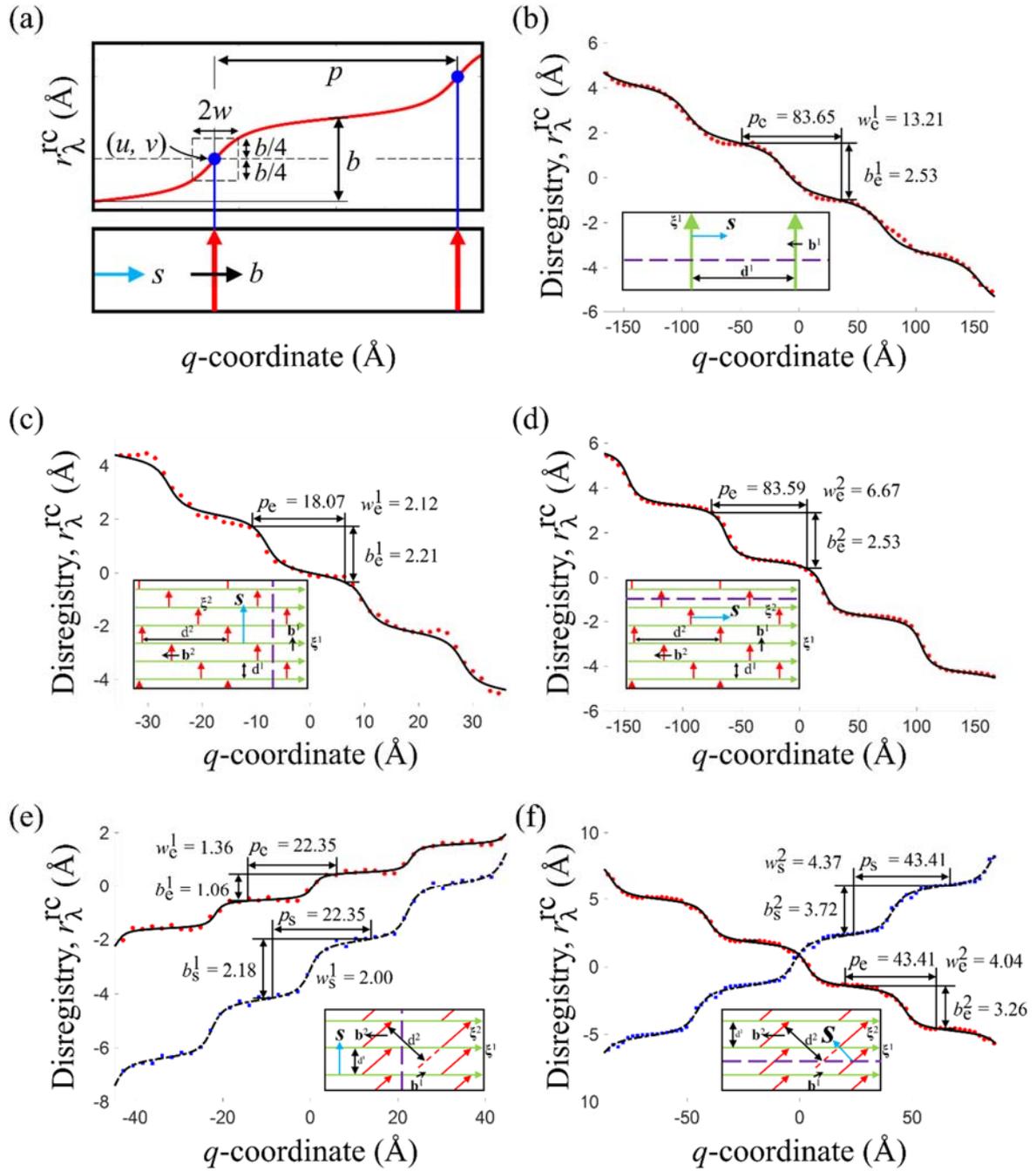

Figure 3. (a) Schematic diagram of disregistry analysis using Equation 1, disregistry along the $q$-coordinate for (b) Isaichev (IS), (c-d) Near Bagaryatsky (Near BA) and (e-f) Near Pitsch-Petch (Near PP) orientation relationship; Purple dashed line indicates the data points used in disregistry analysis for each graph. Black solid and dashed-dotted line represent the fitted function for edge ($r_e^{rc}$) and screw ($r_s^{rc}$) component of disregistry, respectively. Circle and square marker represent the edge ($r_{ij,e}^{rc,atom}$) and screw ($r_{ij,s}^{rc,atom}$) component of disregistry computed from atomistic simulation, respectively.

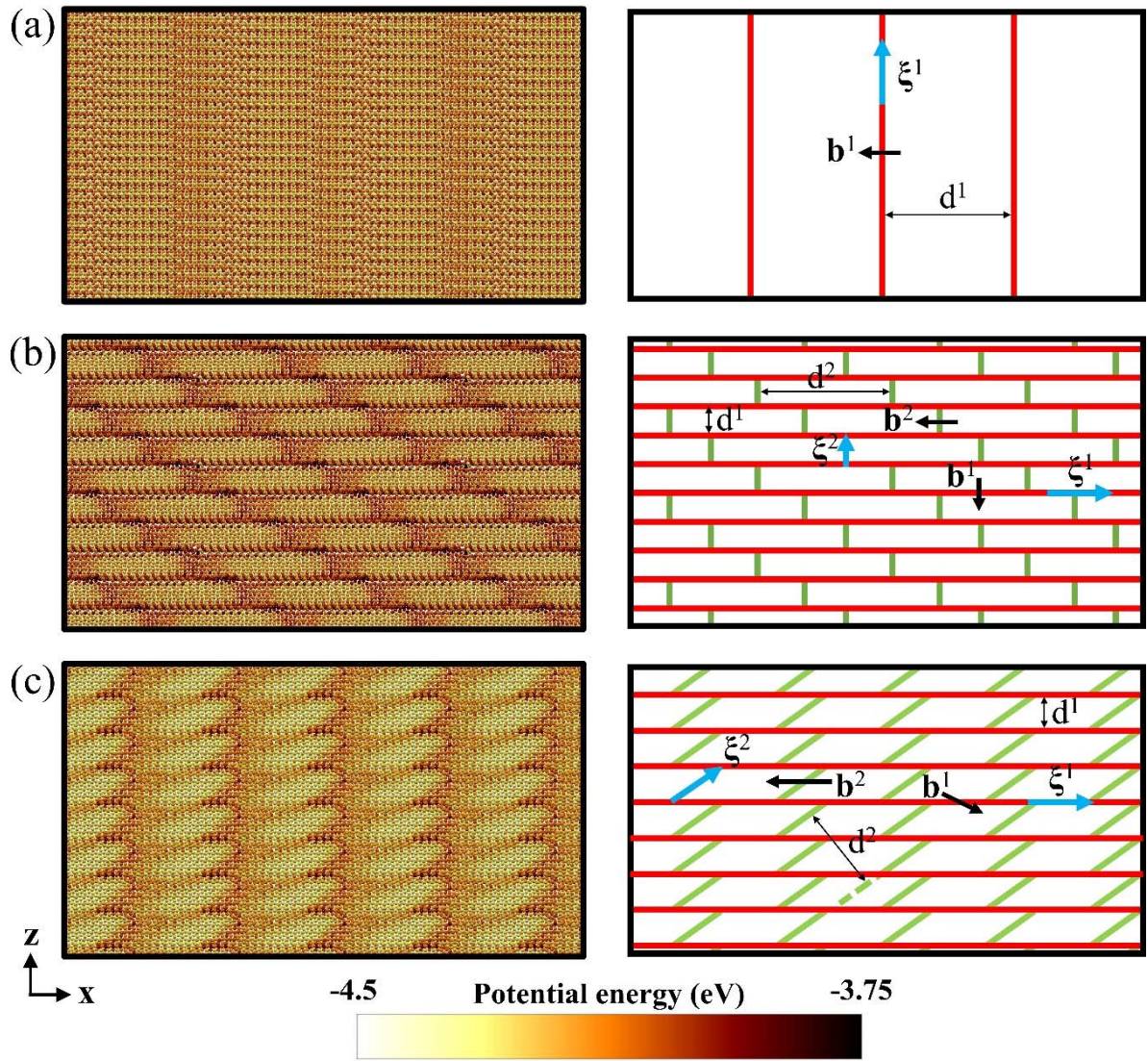

Figure 4. Potential energy maps (left) and idealized misfit dislocation structures (right) of the ferrite/cementite interface for (a) Isaichev (IS), (b) Near Bagaryatsky (Near BA) and (c) Near Pitsch-Petch (Near PP) orientation relationship; The dark region in potential energy map represent not only the local incoherency of the interface but also the core region of the misfit dislocation. For Isaichev orientation relationship, there is small potential energy variation along the x-axis, which implies that the core of misfit dislocation widely spreads along the x-axis.

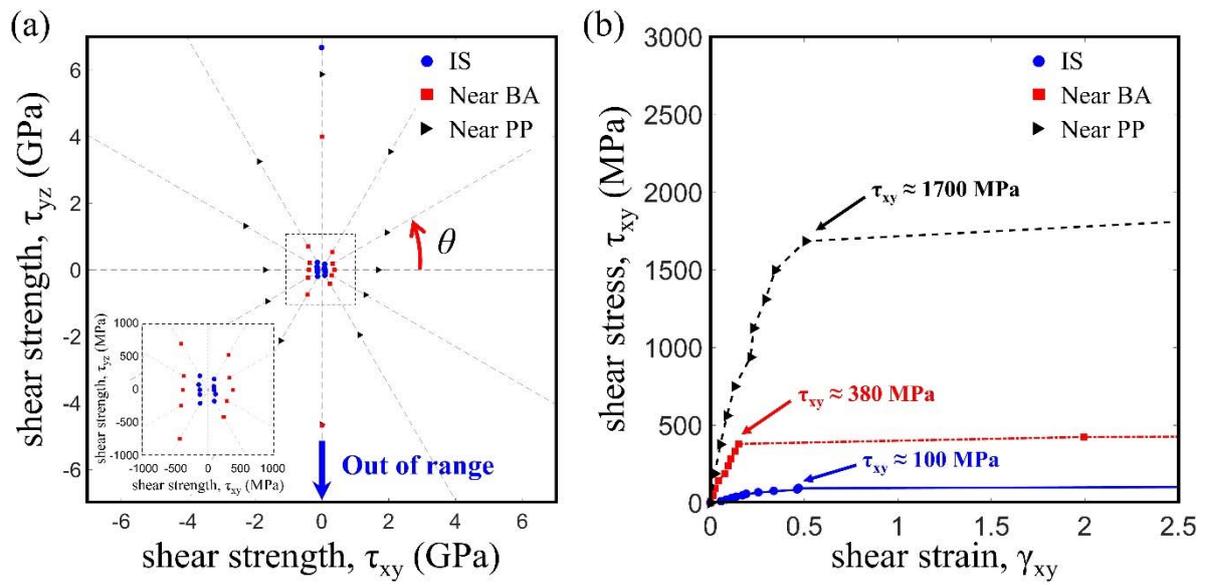

Figure 5. (a) in-plane shear strength map and (b) stress-strain curve for $\theta = 0°$ of the ferrite/cementite bilayer for Isaichev (IS, blue circle), Near Bagaryatsky (Near BA, red square) and Near Pitsch-Petch (Near PP, black triangle) orientation relationships

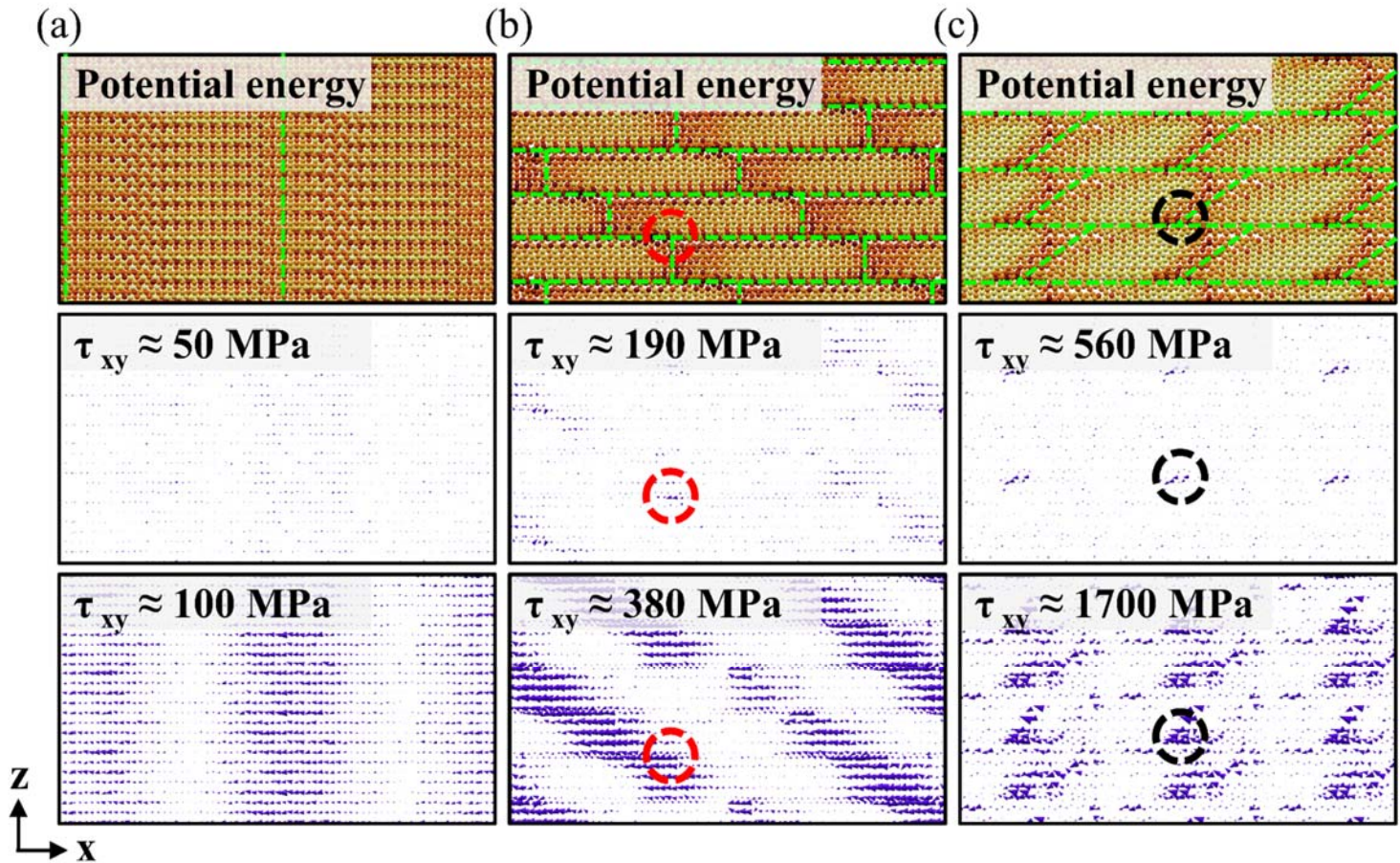

Figure 6. Potential energy maps of initial (relaxed) ferrite/cementite interface and 2D disregistry ($r_{ij}^r$) maps of the ferrite/cementite bilayer for (a) Isaichev (IS), (b) Near Bagaryatsky (Near BA) and (c) Near Pitsch-Petch (Near PP) orientation relationships under low and high in-plane shear stresses for $\theta = 0°$; Red and black dashed circles indicate the junction of the misfit dislocations for Near Bagaryatsky (Near BA) and Near Pitsch-Petch (Near PP) orientation relationships, respectively; The green dashed lines represent the idealized misfit dislocation on the ferrite/cementite interface for each orientation relationship.

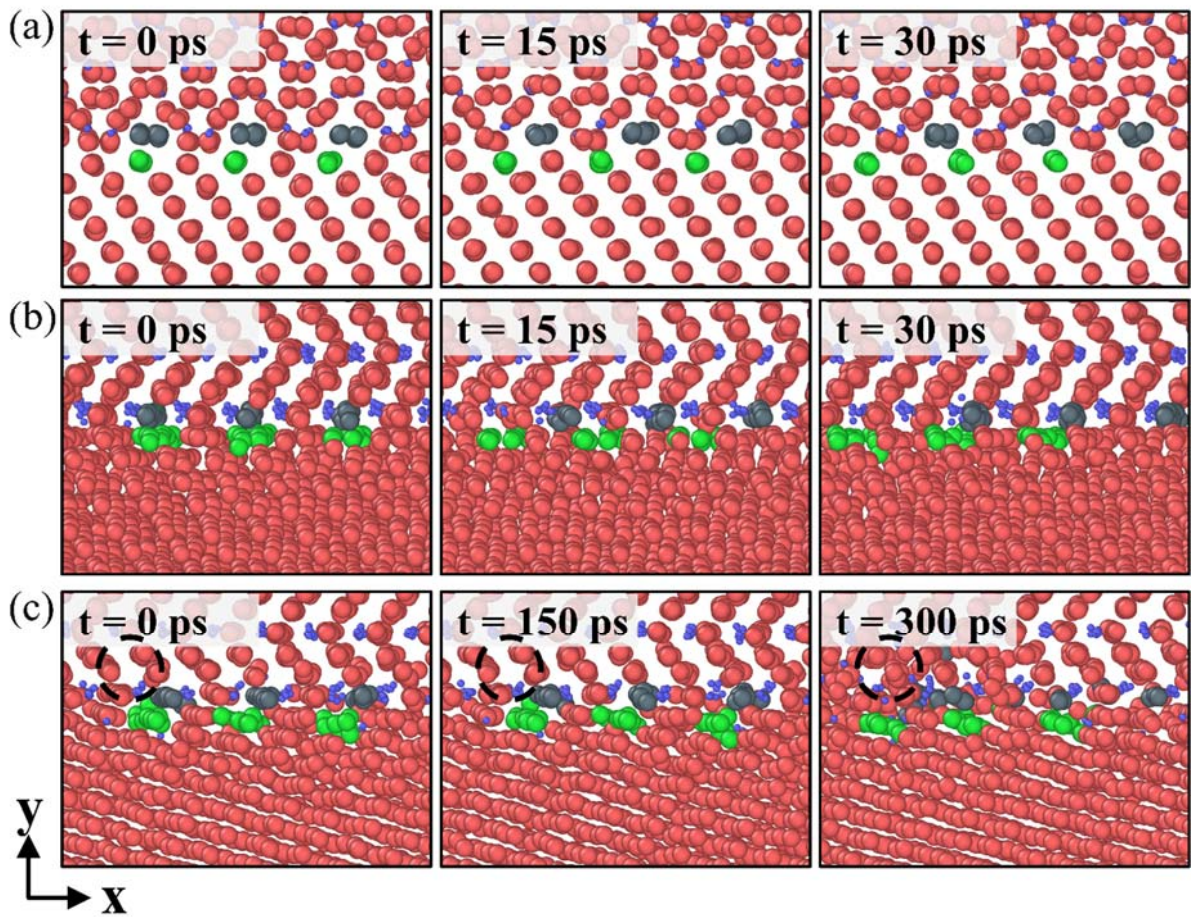

Figure 7. A series of snapshot at the FCI under applied shear stress $\tau_{xy}$ for (a) Isaichev (IS, $\tau_{xy} \approx 110$ MPa), (b) Near Bagaryatsky (Near BA, $\tau_{xy} \approx 570$ MPa) and (c) Near Pitsch-Petch (Near PP, $\tau_{xy} \approx 1870$ MPa) orientation relationships; Pink and blue atoms represent Fe and C atoms, respectively. Gray and green atoms represent the Fe atoms at the interface region of cementite and ferrite, respectively. Black dashed circle in Figure 7-c indicates the fracture region.

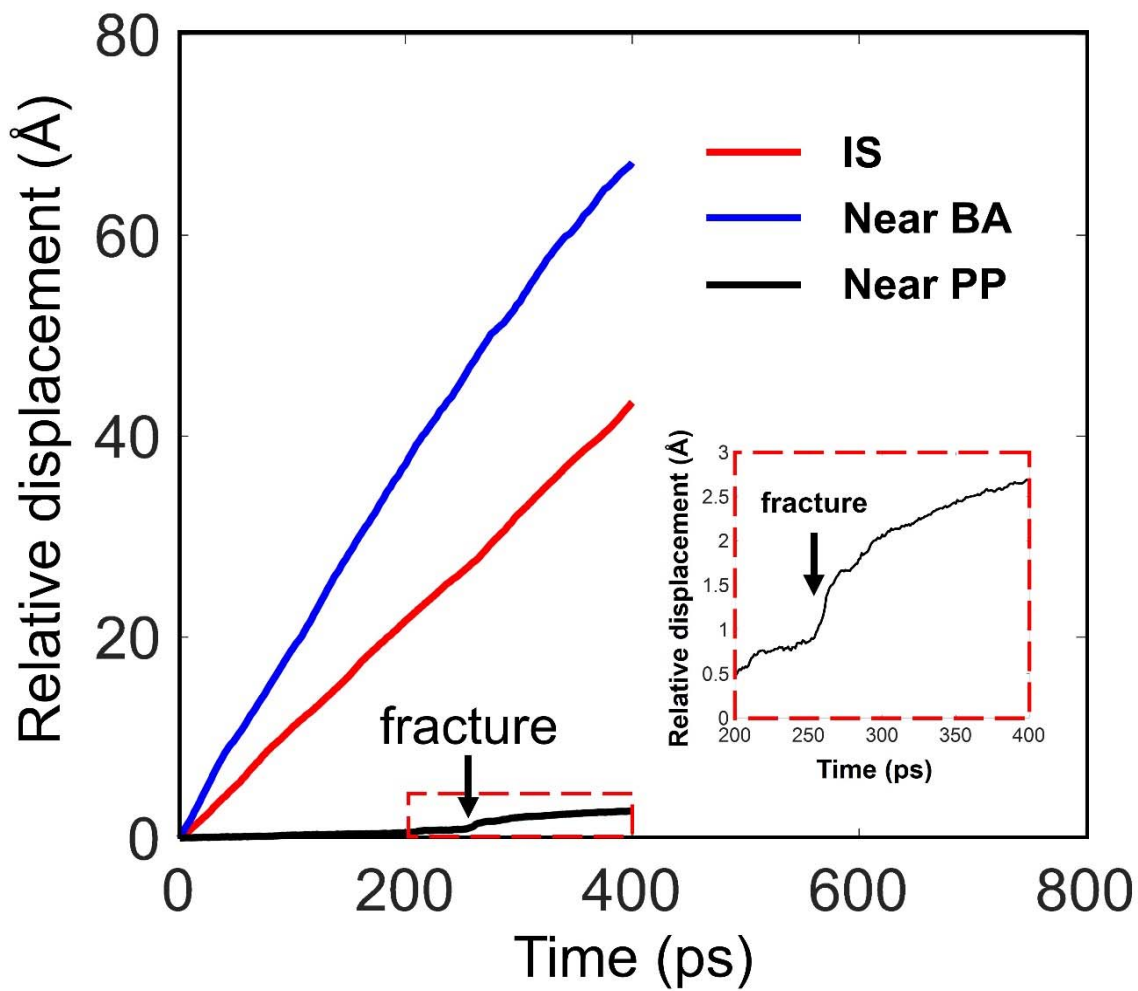

Figure 8. Time versus relative displacement of interface atoms in x-direction for Isaichev (IS), Near Bagaryatsky (Near BA) and Near Pitsch-Petch (Near PP) orientation relationship; We applied constant in-plane shear stress for each orientation relationship for $\theta = 0°$ ($\tau_{xy}$= 110, 570 and 1870 MPa for IS, Near BA and Near PP orientation relationship, respectively). For easy comparison, we set the initial relative displacement as zero right after desired shear stress is applied.

**Supplementary Information**

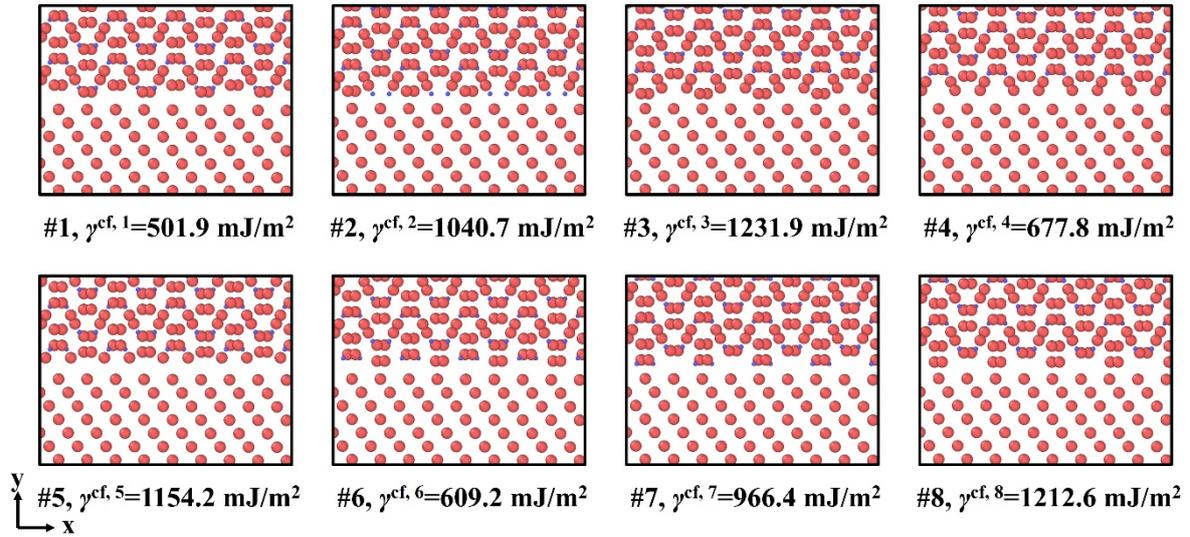

Figure S1. Atomic configurations of un-relaxed interface structures and interface energy for each shuffle plane of Isaichev (IS) orientation relationship; #1 shuffle plane is the most stable interface structure among 8 different shuffle planes.

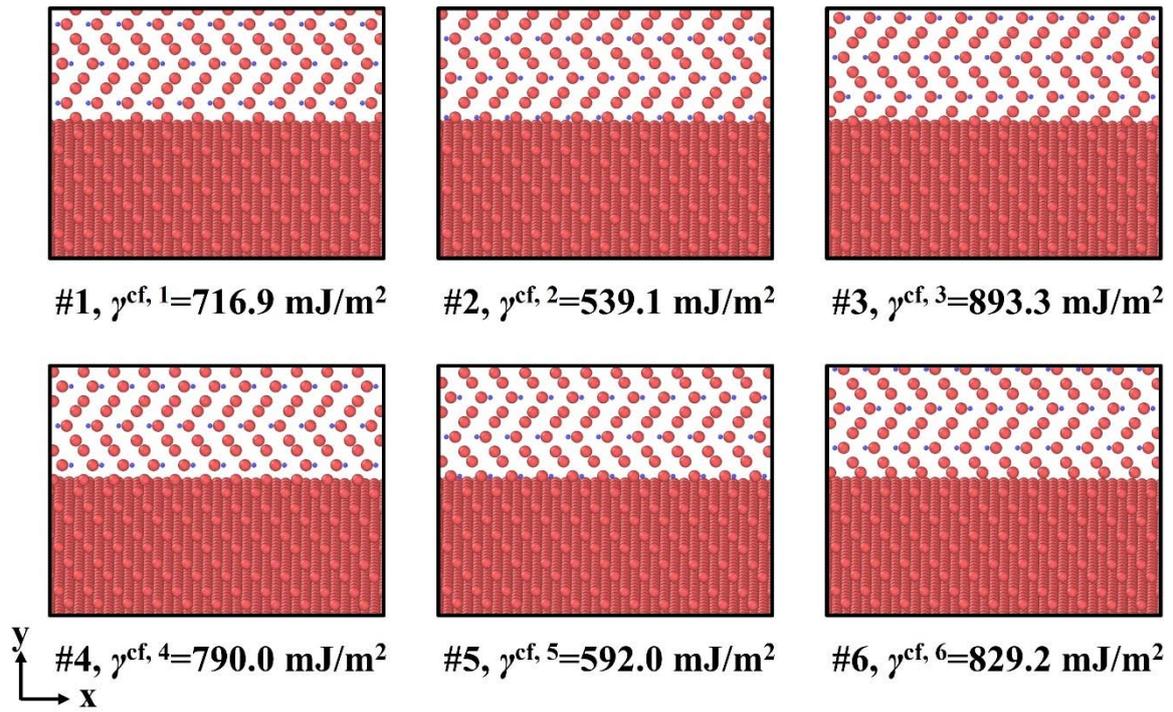

Figure S2. Atomic configuration of un-relaxed interface structure and interface energy for each shuffle plane of Near Bagariatsky (Near BA) orientation relationship; #2 shuffle plane is the most stable interface structure among 6 different shuffle planes.

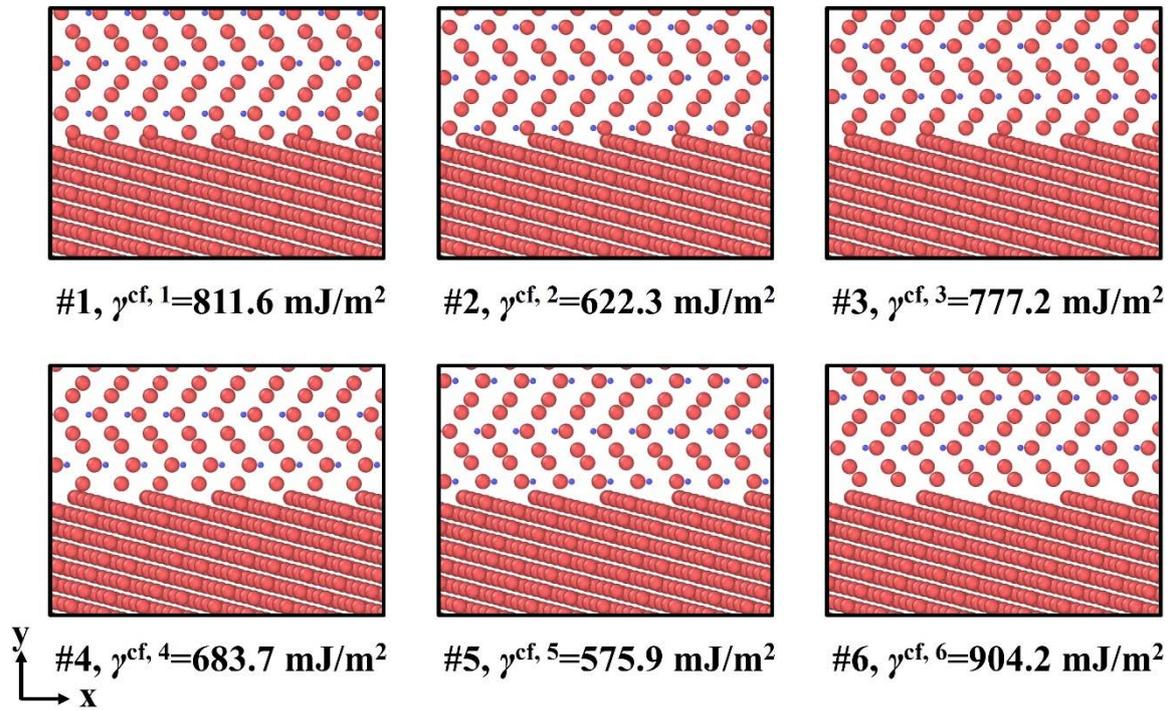

Figure S3. Atomic configuration of un-relaxed interface structure and interface energy for each shuffle plane of Near Pitsch-Petch (Near PP) orientation relationship; #5 shuffle plane is the most stable interface structure among 6 different shuffle planes.

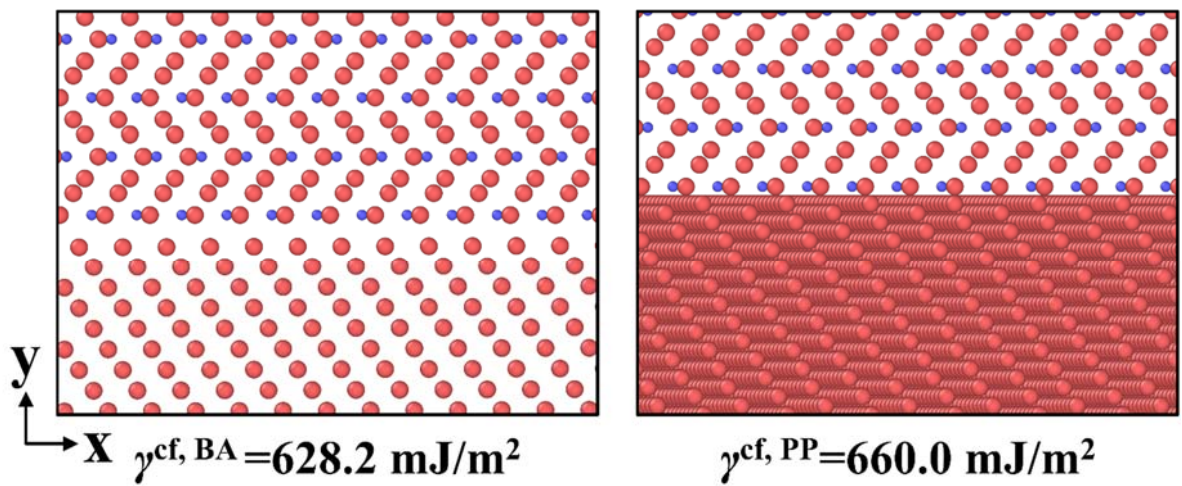

Figure S4. Atomic configurations of un-relaxed interface structures and interface energies for Bagaryatsky (left) and Pitsch-Petch (right) orientation relationship; Each figure represents the most stable interface structure among 6 different shuffle planes for each orientation relationship.

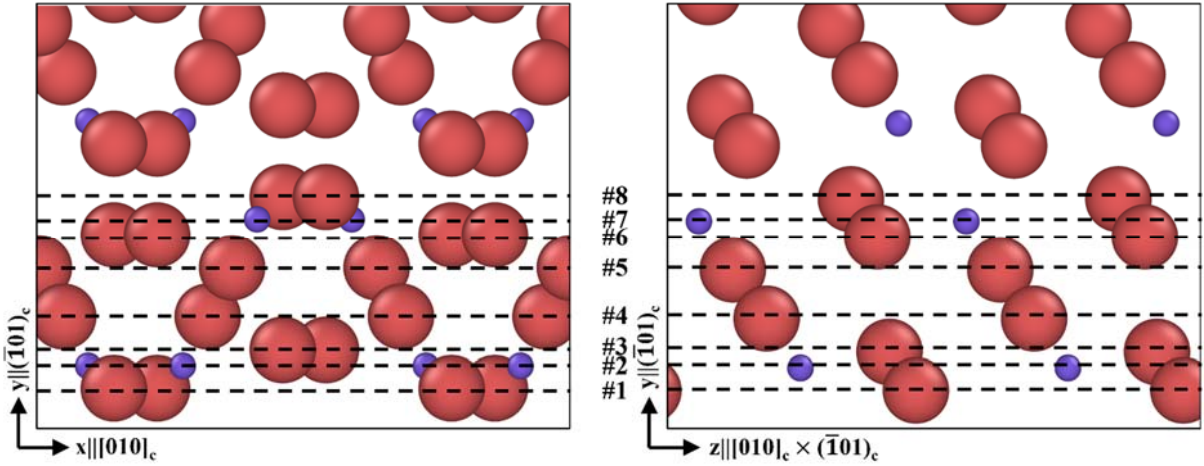

Figure S5. The eight shuffle planes of cementite block for Isaichev (IS) orientation relationship. The five shuffle planes (#1 to #5) are identical to the reported shuffle planes in the reference (Zhou, Zheng et al. 2017) but the three shuffle planes (#6 to #8) were not reported.

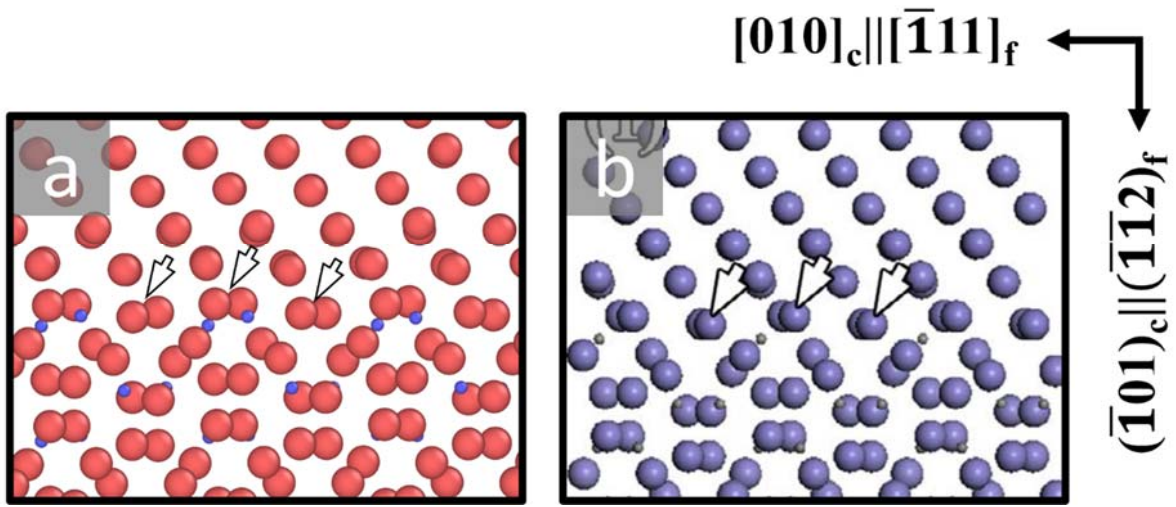

Figure S6. Atomic configurations of ferrite/cementite interface (a) computed by MEAM potential in this study and reported by Zhou et al. (Zhou, Zheng et al. 2017) for Isaichev (IS) orientation relationship; Shuffle plane is identical but distance between Fe atoms indicated by arrows are different.